\documentclass[12pt]{article}
\usepackage{a4}
\usepackage{bbold}
\def\uni{\mathbb{1}}

\def\bea{\begin{eqnarray}}
\def\eea{\end{eqnarray}}
\def\be{\begin{equation}}
\def\ee{\end{equation}}

\def\ld{\lambda^\dagger}
\def\gam{\gamma}

\def\L{\Lambda}
\def\eps{\epsilon}

\def\ldag{\lambda^\dagger}

\def\ss{\scriptstyle}

\newcommand\rf[1]{~(\ref{#1})}
\def\dx{\partial_x}
\def\dxa{\partial_{x_a}}
\def\dag{\dagger}
\def\nn{\nonumber}

\def\OF{|0_{\ss F}\rangle}
\def\OB{|0_{\ss B}\rangle}
\def\OR{|0_{\ss B},0_{\ss F}\rangle}

\def\ttheta{\tilde{\theta}}
\def\la{\langle}

\def\CM{{\rm CoM}}
\def\ra{\rangle}
\def\INT{{\rm Int}}
\def\m{\mu}
\def\n{\nu}
\def\g{\gamma}
\def\a{\alpha}
\def\b{\beta}
\def\ha{{\hat{\alpha}}}
\def\hb{{\hat{\beta}}}
\def\pr{\prime}
\def\ha{{\hat{\alpha}}}

\newcommand{\sect}[1]{\setcounter{equation}{0} \section{#1}}
\renewcommand{\theequation}{\thesection .\arabic{equation}}
\def\lbf{\la 0_B,0_F|}
\def\rbf{|0_B,0_F\ra}
\def\rar{\rightarrow}             
\def\H{{\cal H}}
\def\ldag{\lambda^\dagger}

\def\ss{\scriptstyle}

\def\ld{\l^{\dagger}}
\def\min{|-\ra}
\def\pl{|+\ra}
\def\pmin{|\pm\ra}
\def\minp{|\mp\ra}

\def\l{\lambda}
\newcommand{\NPB}[3]{Nucl.\ Phys. B#1 (#2) #3}

\newcommand{\PRD}[3]{Phys.\ Rev.\ D#1 (#2) #3}
\newcommand{\PLB}[3]{Phys.\ Lett.\ B#1 (#2) #3}

\begin{document}

\thispagestyle{empty}
\begin{flushright}
{\sc\footnotesize hep-th/9710104}\\
{\sc NIKHEF} 97-040
\end{flushright}
\vspace{1cm}
\setcounter{footnote}{0}
\begin{center}
{\LARGE{On the quantum mechanics of $M$(atrix) theory.}
    }\\[14mm]

\sc{
Jan Plefka\footnote{e-mail:
plefka@nikhef.nl} and Andrew Waldron\footnote{e-mail:
waldron@nikhef.nl}}\\[5mm]
{\it NIKHEF, P.O. Box 41882, 1009 DB Amsterdam,\\
The Netherlands}\\[20mm]

{\sc Abstract}\\[2mm]
\end{center}
We present a study
of $M$(atrix) theory from a purely canonical viewpoint.
In particular, we identify free particle
asymptotic states of the model corresponding to the supergraviton
multiplet of eleven dimensional supergravity. These states have a
natural
interpretation as excitations in the flat directions of the matrix model
potential. Furthermore, we provide the split of the matrix model
Hamiltonian
into a free part describing the free propagation of these particle states
along with the interaction Hamiltonian describing their interactions.
Elementary quantum mechanical perturbation theory then yields an effective
potential for these particles as an expansion
in their inverse separation. 
Remarkably we find that the leading
velocity independent terms of the effective potential  
cancel in agreement
with the fact that
there is no force between stationary $D0$ branes. The scheme we present
provides a framework in which one can perturbatively compute the 
$M$(atrix) theory result for the eleven dimensional supergraviton 
$S$ matrix.

\vfill
\leftline{{\sc October 1997}}

\newpage
\setcounter{page}{1}
\sect{Introduction.}

$M$(atrix) theory \cite{BFSS96}
is the conjectured description of $M$ theory in
terms of a quantum mechanical supersymmetric $U(N)$ matrix model~\cite{CH} 
and as 
such has recently been the subject of intense study.
At low energies and large distances $M$ theory, by definition, reduces
to eleven dimensional supergravity~\cite{Witten95}. 
Thus a principal test of the  
conjecture
is the computation of graviton scattering amplitudes
in $M$(atrix) theory followed by a comparison to the supergravity result.
According to Susskind \cite{Susskind97}, equivalence is expected
to hold even for finite $N$, where $M$(atrix) theory is conjectured
to provide the discretized light cone quantization of  $M$ theory.

To date all computations have been based on path integral
quantization and semiclassical expansions thereof,
yielding 
several rather spectacular consistency tests \cite{DKPS97,Ael97,BB97,BBPT97}.
However, although dominant in
many quantum field theoretical contexts, path integrals have
played a subdominant r\^ole in the development of quantum mechanics
for which canonical methods reign. In this paper we study $M$(atrix)
theory from a purely canonical point of view, concentrating on the
two body sector of the theory with each particle carrying one unit of
quantized light cone momentum.
Let us stress that our ultimate goal is the computation of $M$(atrix)
theory supergraviton $S$ matrix elements, rather than to test the well
known equivalence between path integral and canonical methods.

The calculation of scattering amplitudes in gauge theories typically
splits into two steps. The first involves the computation of Greens
functions subject to Ward identities expressing the gauge invariance 
of the theory. These Greens functions are neatly encapsulated in the 
effective action or generating functional of one particle
irreducible diagrams. The perturbative
calculation of the effective action can be
performed efficiently using path integral methods in which the  
gauge is fixed
via the Faddeev Popov procedure~\cite{FP67}. In particular
the background field formalism~\cite{Background} 
provides an elegant framework for
such calculations. The above techniques have already been rather
successfully applied to $M$(atrix) theory and the connection between
semiclassical path integral quantization around 
particular classical backgrounds
and the eikonal approximation has been exploited to extract physical 
results~\cite{DKPS97,Ael97,BB97,BBPT97}.

However, typically one is interested in $S$ matrix elements
for which a second step is required. Namely, the asymptotic
analysis of Lehmann, Syzmanzik and Zimmermann (LSZ)~\cite{LSZ59} 
in which one  
identifies
the asymptotic states of the theory and derives reduction formulae
describing the connection between off-shell Greens functions and physical
matrix elements. A central theme of this work is the
``LSZ formalism'' for $M$(atrix) theory\footnote{It is to be emphasised 
that $M$(atrix) theory is an on-shell theory as regards
eleven dimensional energy momentum, but we are interested in the 
connection between path integral $M$(atrix) theory effective actions and  
the eleven dimensional supergraviton $S$ matrix.}. The bounty is  
obvious, since
such a formalism allows the (perturbative)
computation of the $M$(atrix) theory supergraviton $S$ matrix.

Our primary result is the identification of the $M$(atrix) theory
asymptotic supergraviton particle states. Furthermore, the $M$(atrix) theory 
is a first quantized quantum mechanical theory so that knowledge of these
asymptotic states allows the formulation of
``reduction formulae'' relating 
$S$ matrix elements to covariant gauge path integrals. 
These formulae may be found in our concluding remarks
and their derivation, being relatively standard, is given in the 
appendix.

Since the model is quantum mechanical,
rather than a quantum field theory, one may also directly
compute the $S$ matrix by
applying the techniques
of quantum mechanical scattering theory to our asymptotic states.
This computation amounts to studying $M$(atrix) theory in
the temporal gauge and one suspects that a covariant gauge path integral 
computation along with the reduction of off-shell Greens functions 
to physical amplitudes via an LSZ type analysis involving our 
asymptotic states ultimately provides the most efficient
computational framework. Nonetheless, the first results from 
elementary quantum mechanical perturbation theory are very
encouraging. In particular we find the correct cancellation 
of velocity independent terms in the effective potential at the leading
order of perturbation theory.

Quantum mechanical scattering theory is described by identifying the free
part of the
Hamiltonian $H_0$ whose eigenstates may be interpreted as asymptotic
free particle states with interactions
governed by the interaction Hamiltonian $H_{\rm Int}=H-H_0$.
For the $M$(atrix) theory the asymptotic free particle states should 
correspond to
the eleven dimensional supergraviton multiplet. Although classical
backgrounds corresponding to free particle motion are known and form
the basis of the existing semiclassical path integral quantization  
of $M$(atrix) theory, wavefunctions corresponding to asymptotic
particle states have not yet appeared. In particular the description
of the polarisations of asymptotic particles and the dependence
of scattering amplitudes upon these polarisations remains unclear
in the conventional background field approach, whereas it becomes  
transparent
in the canonical treatment as we shall show.

The construction of such asymptotic
wavefunctions is rendered non-trivial by the presence of constraints
expressing the super Yang-Mills
gauge invariance of the $M$(atrix) theory along with
the interpretation  of spacetime as an asymptotic limit of the theory.
Our work provides the resolution of both these issues.

The main results of this paper are the construction of
wavefunctions describing pairs of free supergravitons along with the free
and interaction Hamiltonians describing their free propagation and
interactions.
We also calculate the Born amplitude along with the leading contribution
from second order perturbation theory. 
A  supersymmetric cancellation between the coefficients of the $1/r^2$
contribution to the effective potential is found  
in agreement with existing two loop path integral calculations~\cite{BB97}.
It is shown that spin interactions give no contribution to the Born amplitude
and that the systematic computation of spin effects is made possible by our
framework.

\subsection{Results and Outline.}

Some aspects of our work are rather technical, 
therefore we have found it useful to include in the
introduction a summary of our findings and explanations of  
how they were obtained. Conventions and detailed derivations 
can be found in the following sections.

\subsubsection*{The model.}

The Hamiltonian of the $M$(atrix) theory is that of $10$d
super Yang-Mills dimensionally reduced to $0+1$ dimensions and
arises in a remarkable way
from two rather disparate viewpoints. On the one hand, it emerges as
a regulating theory of the eleven dimensional supermembrane in light
cone gauge quantization \cite{dWHN88} and on the other hand it is 
the effective Hamiltonian describing the short distance properties 
of $D0$ branes \cite{Polchinski95,Witten96,D0-branes}. Employing the conjecture
of \cite{Susskind97}, the finite $N$ model is to be identified with
the compactification of a null direction of $M$ theory (henceforth
called the $-$ direction). The quantized total momentum of the
$U(N)$ system in this direction
is then given by $P_-=N/R$, where $R$ denotes the compactification
radius. 

We shall be primarily interested in the $U(2)$ theory, studying
the Hilbert space of two supergravitons with momentum $P_-=1/R$ each.
The coordinates and Majorana spinors 
of the transverse nine dimensional space then take values in the adjoint 
representation of $U(2)$, i.e.
\bea
X_\m &=& X^0_\m\,  i\,\uni + X^A_\m\, i\sigma^A \qquad \m=1,\ldots,9 \\
\theta_\alpha &=& \theta_\alpha^0\, i\, \uni + \theta_\alpha^A\, i\sigma^A
\phantom{\sigma}
\qquad \alpha=1,\ldots, 16
\eea
where $\sigma^A$ are the Pauli matrices. We shall often employ a vector
notation for the $SU(2)$ part in which 
$\vec{X}_\m= (X^1_\m,X^2_\m,X^3_\m)\equiv (X^A_\mu)$
and similarly for $\vec{\theta}$. 

The Hamiltonian is then given by
\be
H=H_{\rm CoM}+
R\left(
\frac{1}{2}\vec{P}_\m\cdot\vec{P}_\m
+M^{6}\frac{1}{4}(\vec{X}_\m\times\vec{X}_\m)^2+
M^{3}\frac{i}{2}\vec{X}_\m\cdot
\vec{\theta}\g_\m\times\vec{\theta}
\right)
\label{ham}
\ee
where $H_{\rm CoM}=\frac{1}{2}RP^0_\m P^0_\m$ is the $U(1)$ 
centre of mass Hamiltonian. These two contributions to $H$ are
completely independent and may both be written as a square of a
supersymmetry charge \cite{dWHN88}. 
$M$ denotes the eleven
dimensional Planck mass, which we from now on set to unity. Due to
the linear dependence of the Hamiltonian on $R$ this quantity 
may also be dropped. The explicit $M$ and $R$ dependence
can be reinstated at any stage by a dimensional analysis. 
Note that we are using a real,
symmetric representation of the $SO(9)$ Dirac matrices in which the nine
dimensional charge conjugation matrix is equal to unity. 

The Hamiltonian\rf{ham} is augmented by the Gauss law constraint
\be
\vec{L}=\vec{X}_\m\times\vec{P}_\m-
\frac{i}{2}\vec{\theta}\times\vec{\theta}\quad,\qquad
[L^A,L^B]=i\, \epsilon^{ABC} L^C\label{constraint}
\ee
whose action is required to vanish on physical states.

The task is now to identify the free asymptotic two-particle states of
the Hamiltonian\rf{ham} which describe the on-shell supergraviton
multiplet of eleven dimensional supergravity. This problem manifestly
factorises into a $U(1)$ centre of mass state and an $SU(2)$ invariant
state describing the relative dynamics of the particles.

\subsubsection*{The centre of mass theory.}

The eigenstates of the free $U(1)$ centre of mass matrix theory
are
\be
|k_\mu;h_{\m\n},B_{\m\n\rho},h_{\m\ha}\ra_{_0}=
e^{ik_\m X^0_\m}
|h_{\m\n},B_{\m\n\rho},h_{\m\ha}\ra_{_0}\label{freesoln}
\ee
and possess $SO(9)$ momentum $k_\m$ and on-shell $SO(9)$
supergraviton polarisations\footnote{Note that the polarisation tensors 
$h_{\m\n}$, $B_{\m\n\rho}$ and $h_{\m\ha}$ correspond to {\it physical}
polarisations. The $M$(atrix) theory does away with unphysical
timelike and longitudinal polarisations at the price of manifest
eleven dimensional Lorentz invariance.} $h_{\m\n}$,
$B_{\m\n\rho}$ and $h_{\m\ha}$ (graviton, antisymmetric tensor and  
gravitino,
respectively).
The state $|h_{\m\n},B_{\m\n\rho},h_{\m\ha}\ra_{_0}$  is the
$\underline{44}\oplus\underline{84}\oplus\underline{128}$  
representation of
the centre of mass spinor degrees of freedom.
The construction of this state is carried out in section~\ref{section2}
and allows explicit calculations of the spin dependence of $M$(atrix)
theory supergraviton amplitudes to be carried out.

\subsubsection*{Asymptotic states.}

Relative motions are described in the $M$(atrix) theory by the  
constrained
$SU(2)$ quantum mechanical matrix theory defined above. However, spacetime
is only an asymptotic concept in this theory. In particular diagonal
matrix configurations, i.e., those corresponding to Cartan generators
of $SU(N)$, span flat directions in the matrix model potential and
describe spacetime configurations~\cite{BFSS96}. 
Transverse directions are described
by supersymmetric harmonic oscillator degrees of freedom.

Due to the gauge constraint\rf{constraint}
quantum mechanical wavefunctions must be invariant under $SU(2)$  
rotations
so that there is no preferred Cartan direction. This is not a contradiction
with the identification of spacetime 
spatial degrees of freedom with the diagonal  
Cartan
degrees of freedom of the $M$(atrix) model 
since we only require that the concept of spacetime
emerges in an asymptotic limit.

To find asymptotic states corresponding to supergraviton
(i.e., spacetime) excitations in a gauge invariant way we proceed  
as follows.
Let us suppose we wish to study states describing particles widely
separated in the (say) ninth spatial direction, then we may simply
declare the $SU(2)$ vector $\vec{X}_9$ to be large. The limit
$|\vec{X}_9|=\sqrt{\vec{X}_9\cdot\vec{X}_9}\rightarrow\infty$ is
$SU(2)$ rotation (and therefore gauge) invariant. We search for
asymptotic particle-like solutions in this limit. The solution of
this problem along with the identification of the free and interaction 
Hamiltonians pertaining to their propagation and interactions is  
the subject of
section~\ref{section3}.

To this end it is convenient to employ the (partial) gauge choice in
which one chooses a frame where $\vec{X}_9$ lies along the $z$-axis,
\be
X^1_9=0=X^2_9\, . \label{ed}
\ee
Calling $X_9=(0,0,x)$ and $\vec{X}_a=(Y^1_a,Y^2_a,x_a)$ (with  
$a=1,...,8$) the Hamiltonian in this frame  
includes the
terms\footnote{The spinors $\tilde{\theta}$ are built from
$\theta^1$ and $\theta^2$ by  
complexification and
a ${\rm spin}(9)$ rotation (see equations\rf{comp} and\rf{rotissary}). 
Note that $r^2\equiv x_a x_a+x^2$.}
\bea
H_{\rm V}&=& -\frac{1}{2x}\, (\dx)^2x -\frac{1}{2}\, (\dxa)^2 \\
H_{\rm B}&=& -\frac{1}{2}\, (\frac{\partial}{\partial {Y_a^I}})^2
+ \frac{1}{2}\, r^2\, Y_a^I\, Y_a^I \label{tom}\\
H_{\rm F}&=& r\, \tilde{\theta}^\dag\, \gam_9\, \tilde{\theta}\, .
\eea
The sum of the  Hamiltonians $H_{\rm B}$ and
$H_{\rm F}$ is that of a supersymmetric
harmonic oscillator with frequency $r$ and describes excitations 
transverse to the flat directions. Particle motions in the flat  
directions
correspond to the Hamiltonian $H_{\rm V}$ whereby we interpret
the Cartan variables $x_\m=(x_a,x)$ asymptotically 
as the $SO(9)$ space coordinates.

The Hilbert space may be treated as a ``product'' of transverse  
superoscillator
degrees of freedom and Cartan wavefunctions depending on $x_\m$ and the
third component of $\vec{\theta}$ via the identity
\bea
H=\sum_{m,n}|m\ra\, \la m|H|n\ra\, \la n|
\eea
where $\{|n\ra\}$ denote the complete set of eigenstates of $H_{\rm  
B}$ and
$H_{\rm F}$. Since the frequency $r$ of the superoscillators is  
coordinate
dependent, operators $\partial/\partial x_\m$ do not commute with $|n\ra$
so that this ``product'' is not direct.
This construction allows us to study an ``effective''  
Hamiltonian
$H_{mn}=\la m|H|n\ra$ for the Cartan degrees of freedom pertaining to
asymptotic spacetime. 
Similarly, note that $H_{mn}$ is a differential operator 
in the Cartan variables. 
In particular the free Hamiltonian is given by
the diagonal terms\footnote{We subtract terms $c_n/r^2$ to ensure  
the correct
asymptotic behaviour of the interaction Hamiltonian.
A detailed explanation of this point may be found in section 
\ref{section3}.}
\be
H_0=\sum_n \, |n\rangle\, \langle n|\left( H_{\rm V}+H_{\rm B}+H_{\rm F} 
\, -\,\frac{c_n}{r^2}\right) |n\rangle\,
\langle n|.\label{free}
\ee
Since supersymmetric harmonic oscillator zero point energies vanish,
eigenstates of\rf{free} are
\be
|k_\m;h_{\m\n},B_{\m\n\rho},h_{\m\ha}\ra=\frac{1}{x}e^{ik_\m x_\m}
|h_{\m\n},B_{\m\n\rho},h_{\m\ha}\ra\otimes|0_B,0_F\ra\label{solutions}
\ee
where $|0_B,0_F\ra$ is the supersymmetric harmonic oscillator vacuum.
These states satisfy the correct free particle dispersion relation
\be
H_0|k_\m;h_{\m\n},B_{\m\n\rho},h_{\m\ha}\ra
=\frac{1}{2}k_\m k_\m\, |k_\m;h_{\m\n},B_{\m\n\rho},h_{\m\ha}\ra\ .
\ee
Here, the supergraviton polarisation multiplet
$|h_{\m\n},B_{\m\n\rho},h_{\m\ha}\ra$ is built from the
$\underline{44}\oplus\underline{84}\oplus\underline{128}$  
representation of $\theta^3$.

What are the properties of these states? Firstly, it is to be stressed
that $k_\m$ and $x_\m$ are not $SO(9)$ vectors since manifestly the gauge
choice\rf{ed} breaks $SO(9)$ covariance (the same statement holds for  
supersymmetry).
However, the solutions\rf{solutions}
may easily be written in the original variables by undoing the gauge 
fixing procedure. 
In the limit $|\vec{X}_9|\rightarrow\infty$ one finds
that $k_\m$, $h_{\m\n}$, $B_{\m\n\rho}$ and $h_{\m\ha}$ transform  
correctly
under both $SO(9)$ and supersymmetry. I.e., we have found states which
asymptotically have the required quantum numbers to describe  
supergravitons.

Furthermore in the limit $|\vec{X}_9|\rightarrow\infty$, we find
$H\rightarrow H_0$ and that our particle states
have eigenvalues $k_\m$ of the momentum operator
$|\vec{X}_9|^{-1}\vec{X}_9\cdot\vec{P}_\m$.

Therefore, upon taking the direct product of an asymptotic  
state\rf{solutions}
with a centre of mass eigenstate\rf{freesoln}, one obtains a state
describing a pair of supergravitons widely separated in the ninth
spatial direction whose interactions are governed by the interaction 
Hamiltonian $H_{\rm Int}=H-H_{\rm CoM}-H_0$.

\subsubsection*{Born amplitude.}

The leading contribution to the quantum mechanical scattering amplitude
is the Born amplitude
\be\la k^\pr_\m;h^\pr_{\m\n},B^\pr_{\m\n\rho},h^\pr_{\m\ha}|
\, H_{\rm Int}\, |
k_\m;h_{\m\n},B_{\m\n\rho},h_{\m\ha}\ra
\ee
which yields a nine dimensional Fourier transform of an
effective potential \linebreak \mbox{$V(x_\m$, $\partial_\m)$} in which the 
superoscillator degrees are ``integrated out''.
In section~\ref{section4} we show that the result for $V$ is
\bea
V&=&\frac{16}{r^2} +\frac{(r^2-x^2)^2}
{2\, x^2\, r}
 -\frac{r^2-x^2}{2\, x^2\, r^2}
+\frac{19r^2+x^2}{2x^2\, r^5}
 \nn\\&&
+\frac{x^a}{2\, x^2\, r^3}\, \dxa-\frac{1}{2\, x^2\, r}\, (\dxa)^2.
\label{one}
\eea
The first term was also obtained in~\cite{SS} in a different setting.
The result\rf{one} is encouraging, firstly, since
the leading $1/r^2$ term is $SO(9)$ invariant in  
accordance with
our argument that $SO(9)$ invariance should be recovered in the
large $x=|\vec{X_9}|$ limit. Secondly, even though we have not yet recovered
the revered $v^4/r^7$ potential for $D0$ particles,
the $1/r^2$ term is, in fact, the leading\footnote{Actually a tree level 
contribution proportional to $r$ is also allowed by dimensionality but is
clearly absent.} term allowed on dimensional grounds
in a loop expansion of the effective action of the original 
super Yang-Mills model. However, in explicit calculations \cite{BB97} such a
velocity independent $1/r^2$ term has
been shown to be absent. Fortunately, it is not hard to see that second 
(but not higher) order 
perturbation theory can also yield a $1/r^2$ contribution and in 
section~\ref{section4} we show that the result
is $-16/r^2$. This supersymmetric cancellation yields a strong test of 
our proposal.

\sect{The centre of mass $U(1)$ matrix theory.}
\label{section2}

The $U(N)$ matrix theory may be decomposed into a free $U(1)$ supersymmetric
matrix theory
representing the centre of mass motion of the system and 
a $SU(2)$ theory describing relative motions. The two systems are
independent and in this section we give the complete description of 
the $U(1)$ part. 

The $U(1)$ Hamiltonian is given by
\be
H_{\scriptscriptstyle \rm CoM}=
\frac{1}{2}P_\mu^0 P_\mu^0, \quad \mu=1,\ldots,9\label{Ham}
\ee 
and there is no constraint since the structure constants of the $U(1)$
gauge group vanish. The Hamiltonian acts in a phase space spanned by 
the real variables $(X_\mu^0,P_\mu^0)$ and the real sixteen component
$SO(9)$ spinors $\theta^0$ where
\bea
[P_\mu^0,X_\nu^0]&=&-i\delta_{\mu\nu}\\
\{\theta^0_{\hat{\alpha}},\theta^0_{\hat{\beta}}\}&=&
                                      \delta_{\hat{\alpha}\hat{\beta}}, 
\quad {\hat{\alpha},\hat{\beta}}=1,\ldots,16.\label{fermions}
\eea
The Hamiltonian\rf{Ham} is the square of the real, sixteen component
supersymmetry generator
\bea
Q_{\ha} &=& P^0_\mu \left(\gamma_\mu\theta\right)_\ha \label{Q}\\
\{ Q_\ha ,Q_\hb \}
&=&
2H_\CM\delta_{\ha\hb}.\label{alg}
\eea
The $SO(9)$ Dirac matrices $\g_\m$ are symmetric,  satisfy 
$\{\g_\m,\g_\n\}=2\delta_{\m\n}$ and the nine dimensional charge 
conjugation matrix is taken to be unity.

The eigenstates of\rf{Ham} are simply
\be
|k_\mu;h_{\m\n},B_{\m\n\rho},h_{\m\ha}\ra_{_0}=
e^{ik_\m X^0_\m}|h_{\m\n},B_{\m\n\rho},h_{\m\ha}\ra_{_0}
\ee
and are parametrised by the $SO(9)$ momentum vector $k_\m$
and the on-shell $SO(9)$ graviton, antisymmetric tensor and gravitino
polarisation tensors  ($h_{\m\n}$, $B_{\m\n\rho}$, $h_{\m\ha}$,
respectively) of the eleven dimensional supergraviton 
multiplet. It is important to note that the tensors
$h_{\m\n}$ and $h_{\m\ha}$ are traceless and gamma-traceless, respectively.
The state $|h_{\m\n},B_{\m\n\rho},h_{\m\ha}\ra_{_0}$  is the 
$\underline{44}\oplus\underline{84}\oplus\underline{128}$ representation
of the algebra\rf{fermions} and the rest of this section is devoted
to the explicit construction of this representation along with explicit
realizations of the supersymmetry and $SO(9)$ Lorentz transformations of
these states.

In order to define the fermionic vacuum and creation and annihilation operators
we perform a decomposition of the $SO(9)$ Lorentz algebra with
respect to an $SO(7)\otimes U(1)$ subgroup~\cite{dWHN88}. 
This is done as follows.
Firstly split vector 
indices\footnote{Our index conventions are as follows. Nine dimensional
vector indices are given by $\m,\n,\ldots$ . Vector indices $a,b,\ldots$ 
stand for $1,\ldots,8$ whereas $m,n,\ldots$ denote the values $1,\ldots,7$. 
For spinor indices, sixteen dimensional $SO(9)$ indices are denoted by 
$\ha,\hb,\ldots$ and eight dimensional $SO(7)$ indices are given by 
$\a,\b,\ldots$ .}
$\m=(1,\ldots,9)$
as $(m=1,\ldots,7;8,9)$ so that an $SO(9)$ vector $V_\m$ may be
rewritten as $(V_m,V,V^*)$ where $V=V_8+iV_9$ and 
$V^*=V_8-iV_9$. The parameters $\L_{\m\n}$ of an $SO(9)$ Lorentz transformation
decompose with respect to $SO(7)\otimes U(1)$ into $\L_{mn}$ and $\L_{89}$
corresponding to $SO(7)$ and $U(1)$ transformations, respectively, and
the remaining parameters may be written in the $SO(7)\otimes U(1)$
covariant form $l_m=\L_{m8}+i\L_{m9}$ and $l_m^*=\L_{m8}-i\L_{m9}$.
The $SO(9)$ transformation of a vector is then given by
\bea
V_m&\rightarrow& V_m+\L_{mn}V_n+\frac{1}{2}(l_mV^*+l^*_mV)\\
V&\rightarrow& V-i\L_{89}V-l_mV_m\\
V^*&\rightarrow& V^*+i\L_{89}V^*-l^*_mV_m^*
\eea
For an $SO(9)$ spinor the same decomposition is made by complexifying,
in particular, for the canonical spinor variables we have  
\bea
\l=\frac{\theta^0_++i\theta^0_-}{\sqrt{2}}\, , \,
\l^\dagger=\frac{\theta^0_+-i\theta^0_-}{\sqrt{2}}\label{napoleon},
\eea
where the subscript $\pm$ denotes projection by $(1\pm\g_9)/2$.
The $SO(9)$ transformations then read\footnote{We employ the following
representation for the $SO(9)$ Dirac matrices \be
\g_\m=
\left\{
\left(
\begin{array}{c|c}
0&\g_m\\\hline-\g_m&0
\end{array}
\right),
\left(
\begin{array}{c|c}
0&I\\\hline I&0
\end{array}
\right),
\left(
\begin{array}{c|c}
I&0\\\hline 0&-I
\end{array}
\right)
\right\}\label{Dirac}
\ee
where the real, antisymmetric $SO(7)$ Dirac matrices satisfy 
$\{\g_m,\g_n\}=-2\delta_{mn}$.}
\bea
\l&\rightarrow&\l-\frac{1}{4}\L_{mn}\g_{mn}\l+\frac{i}{2}\L_{89}\l
+\frac{1}{2}l_m^*\g_m\l^\dagger\label{saint}\\
\l^\dagger&\rightarrow&\l^\dagger
-\frac{1}{4}\L_{mn}\g_{mn}\l^\dagger-\frac{i}{2}\L_{89}\l^\dagger
+\frac{1}{2}l_m\g_m\l\label{martyr}
\eea
The canonical anticommutation relations are now
\be
\{\l_\a,\l^\dagger_\b\}=\delta_{\a\b}\, , \quad \a,\b=1,\ldots,8.
\ee
and we define the fermionic vacuum $|-\ra$ by
\be
\l|-\ra=0.\label{vac}
\ee
We denote the completely filled state by $\pl=\ldag_1\ldots\ldag_8\min$.

The most general state is then some linear combination of 256 states of the 
form
\be
\sum_{i=0}^{8}H_{\a_1\cdots\a_i}\l^\dagger_{\a_1}\cdots\l^\dagger_{\a_i}|-\ra
\, .\label{form}
\ee
It is now a simple exercise to extend the $SO(7)\otimes U(1)$
decomposition to the
polarisation tensors $h_{\m\n}$, $B_{\m\n\rho}$ and $h_{\m\ha}$ and then
identify combinations of states in\rf{form} with the same $SO(7)\otimes
U(1)$ transformation properties.
One finds then the following expansion for the supergraviton
polarisation state
\bea
|h_{\m\n},B_{\m\n\rho},h_{\m\ha}\ra_{_0}&=&h\min+\frac{1}{4}h_{m}\min_m
			+\frac{1}{16}h_{mn}\pmin_{mn}
			+\frac{1}{4}h^*_m\pl_m+h^*\pl\nn\\
			&&\hspace{-.8cm}-\frac{\sqrt{3}\, i}{8}
                 \left(\frac{}{}\!B_{mn}\min_{mn}
			+\frac{i}{6}B_m\pmin_m
                      +\frac{1}{6}B_{mnp}\pmin_{mnp}
          -B^*_{mn}\pl_{mn}\right)\nn\\
&&\frac{i}{\sqrt{2}}
\left(h_\a\min_\a-\frac{1}{2}h_{m\a}\min_{m\a}+\frac{1}{2}h^*_{m\a}\pl_{m\a}
  -h_\a^*\pl_\a\right)
\ .\nn\\&&\label{result}
\eea
The states in\rf{result} are defined in table~\ref{table}. 
Note that we have, without loss of generality, assigned the vacuum the
$U(1)$ weight $2$ so that $|-\ra\rightarrow|-\ra+2i\L_{89}|-\ra$.
The less obvious
$SO(7)\otimes U(1)$ decompositions of the supergraviton
polarisation tensors are defined 
below
\bea
h_m&=&h_{m8}+ih_{m9}\\
h&=&\frac{h_{88}-h_{99}}{2}+ih_{89}\\
B_{mn}&=&B_{mn8}+iB_{mn9}\\ 
B_m&=&B_{m89}\\
h_\a&=&\frac{\left(h_{8+}+ih_{9+}-i[h_{8-}+ih_{9-}]
\right)_{^{\scriptstyle\a}}}{\sqrt{2}}\label{jack}\\
h_{m\a}&=&\frac{\left(h_{m+}-ih_{m-}
\right)_{^{\scriptstyle\a}}}{\sqrt{2}}\ .
\eea 

\begin{table}
\bea
\begin{array}{|clcl|c|c|}
\hline
\, &\mbox{State}&&&\mbox{$U(1)$ weight}&SO(7)\\
\hline\hline&&&&&\\{}&
\min&&&2&\mbox{scalar}\\&&&&&\\&\min_\a&=&\ldag_\a\min&3/2&
\mbox{spinor}\\&&&&&\\{}&
\min_m&=&(\ld \g_m \ld)\min& 1& \mbox{vector}
\\&&&&& \\{}&
\min_{mn}&=&(\ld \g_{mn}\ld)\min&
                        1& \mbox{a/symmetric tensor}\\&&&&& 
\\& \min_{m\a}&=&\ldag_\a(\ldag\g_m\ldag)\min&1/2&\mbox{vector-spinor}\\&&&&&
\\{}&
\minp_{mn}&=&(\ld\g_m\ld)(\ld\g_n\ld)\min&
  			 0& \mbox{symmetric tensor}\\{} &
\hspace{.3cm}{\scriptscriptstyle |\hspace{.5mm} |} &&&& \\{}&
\pmin_{mn}&=&(\l\g_m\l)(\l\g_n\l)\pl&
  			 0& \mbox{symmetric tensor}\\& &&&&\\&
\minp_{mnp}&=&(\ld\g_{[mn}\ld)(\ld\g_{p]}\ld)\min&
			0& \mbox{a/symmetric tensor}\\{}&
\hspace{.3cm}{\scriptscriptstyle |\hspace{.5mm} |} &&&& \\{}&
\hspace{-.32cm}
-\pmin_{mnp}&=&-(\l\g_{[mn}\l)(\l\g_{p]}\l)\pl&
			 0& \mbox{a/symmetric tensor}\\& &&&&\\{}&
\minp_{m}&=&(\ld\g_{nm}\ld)(\ld\g_{n}\ld)\min&
			 0& \mbox{vector}\\{}& 
\hspace{.3cm}{\scriptscriptstyle |\hspace{.5mm} |} &&&& \\{}&
\pmin_{m}&=&(\l\g_{nm}\l)(\l\g_{n}\l)\pl&
			 0& \mbox{vector}\\&&&&& \\{}&
\pl_{m\a}&=&\l_\a(\l\g_{m}\l)\pl&-1/2&\mbox{vector-spinor}
\\& &&&&
\\&\pl_{mn}&=&(\l \g_{mn}\l)\pl&
                        -1& \mbox{a/symmetric tensor} \\&&&&&\\{}&
\pl_m&=&(\l \g_m \l)\pl&-1& \mbox{vector}\\&&&&&\\&\pl_\a&=&\l_\a\pl&-3/2&
\mbox{spinor}\\&&&&&\\{}&
\pl&&&-2&\mbox{scalar}
\\&&&&&\\
\hline
\end{array}
\nonumber
\eea
\caption{States transforming covariantly with respect to $SO(7)\otimes U(1)$. 
Note that $(\ldag\g_m\ldag)\equiv\ldag_\a(\g_m)_{\a\b}\ldag_\b$. 
\label{table}}
\end{table}
The relative coefficient of each term in\rf{result} is not fixed by
$SO(7)\otimes U(1)$ invariance alone. Firstly 
note\footnote{To this end one needs to use the equalities of the states  
$\pmin_{mn}=\minp_{mn}$, $\pmin_{mnp}=-\minp_{mnp}$ and 
$\pmin_{m}=\minp_m$ which follow, respectively, 
from the identities
\bea
(\g_m)_{[\a\b}(\g_n)_{\gamma\delta]}&=&
\frac{1}{4!}
\epsilon_{\a\b\gamma\delta\a^\prime\b^\prime\gamma^\prime\delta^\prime}
(\g_{m})_{\a^\prime\b^\prime}(\g_{n})_{\gamma^\prime\delta^\prime}\nn\\
(\g_{[mn})_{[\a\b}(\g_{p]})_{\gamma\delta]}&=&
-\frac{1}{4!}
\epsilon_{\a\b\gamma\delta\a^\prime\b^\prime\gamma^\prime\delta^\prime}
(\g_{[mn})_{\a^\prime\b^\prime}(\g_{p]})_{\gamma^\prime\delta^\prime}\nn\\
(\g_{nm})_{[\a\b}(\g_n)_{\gamma\delta]}&=&
\frac{1}{4!}
\epsilon_{\a\b\gamma\delta\a^\prime\b^\prime\gamma^\prime\delta^\prime}
(\g_{nm})_{\a^\prime\b^\prime}(\g_n)_{\gamma^\prime\delta^\prime}\nn
\eea
We (anti)symmetrise with unit weight.} 
that\rf{result} is
real with respect to complex conjugation where $\min^*=\pl$.
Having fixed reality, one must employ covariance with respect to the remaining
$SO(9)$ transformations (those with parameters $l_m$ and $l^*_m$)
and supersymmetry to fix the coefficients as given in\rf{result}. From 
the definition\rf{vac} and the transformation law\rf{saint},
one deduces that the vacuum is not inert under the leftover $SO(9)$ 
transformations but rather
\be
\min\rightarrow\min-\frac{1}{4}l_m^*\min_m\label{fervac}
\ee
and similarly
\be
\pl\rightarrow\pl-\frac{1}{4}l_m\pl_m\label{fervac2} .
\ee 
Then using the transformation laws\rf{saint}, (\ref{martyr}), (\ref{fervac})
and\rf{fervac2} 
the transformation laws of the states in table~\ref{table}  
may be calculated. So long as one remembers the tracelessness conditions
on the graviton and gravitino it is not hard to check that the
transformation laws of the states induce the correct $SO(9)$ transformation 
laws for the supergraviton polarisation tensors.

The supersymmetry transformations of the states in table~\ref{table}
are obtained by acting with the explicit supersymmetry charge\rf{Q}
which may be written in $SO(7)\otimes U(1)$ covariant language as
\be
i\Lambda_{\ha} Q_{\ha}=k_m(l^*\g_m\l-l\g_m\lambda^\dagger)+
kl\l-k^*l^*\l^\dagger,
\label{Ql}
\ee
where we have replaced the operator $P^0_\m$ by its eigenvalue $k_\m=
(k_m,k,k^*)$. Note that $l$ and $l^*$ are defined (in terms of
$\Lambda$) analogously 
to $\lambda$ and $\lambda^\dagger$, respectively, in\rf{napoleon}. 
It is easy to check that the algebra is, as required
\be
[i\L_1Q,i\L_2Q]=\frac{1}{2}k_\m k_\m(\L_1)_{\ha}
(\L_2)_{\hb}(2\delta_{\ha\hb}).
\label{algebra}
\ee 
Again it is a simple exercise to verify that the action of the Hermitean
generator $i\L_\ha Q_\ha$ in\rf{Ql} on the state\rf{result} 
induces the correct supersymmetry 
transformations of the supergraviton polarisation tensors 
\bea
h_{\m\n}&\rightarrow&h_{\m\n}+
\frac{i}{\sqrt{2}}\,\L_{\ha} k_\rho\left(\g_{\rho}\g_{(\m}h_{\n)}\right)_{{}^{
{}^{\scriptstyle{\ha}}}}
\label{hmn}\\
B_{\m\n\rho}\hspace{-.25cm}&\rightarrow&B_{\m\n\rho}+
i\sqrt{\frac{3}{8}}\,\L_{\ha} k_\sigma
\left(\g_\sigma\g_{[\m\n}h_{\rho]}\right)_{{}^{{}^{\scriptstyle\ha}}}\\
h_{\m\ha}&\rightarrow&h_{\m\ha}
+\frac{i}{\sqrt{2}}\, k_{\rho}h_{\sigma\m}
\left(\g_{\sigma}\g_{\rho}\L\right)_{\ha}
+\frac{i}{6\sqrt{6}}\, B_{\rho\sigma\eta}k_{\kappa}
\left(\g_\m\g_{\rho\sigma\eta}\g_\kappa\L\right)_\ha\nn\\&& \hspace{1.6cm}
-i\sqrt{\frac{3}{8}}\, B_{\m\rho\sigma}k_\eta
\left(\g_{\rho\sigma}\g_\eta\L\right)_\ha\, .
\label{hm}
\eea
One may check that the right hand sides of equations\rf{hmn} and\rf{hm}
are traceless and gamma-traceless, respectively. Furthermore, the commutator
of two supersymmetry transformations, as given in\rf{hmn}-(\ref{hm}), satisfy 
the algebra\rf{algebra} (this result, of course, 
is ensured since we have an explicit operator
representation\rf{Ql} of the algebra\rf{algebra}). 

Finally, for the sake of physics,
we are interested in inner products of these supergraviton polarisation
states. Defining $\la-|-\ra=\{|-\ra\}^\dagger|-\ra=1$
where $\la-|\l_\a=0$, we find
\be
\langle h_{\m\n}^\prime,B^\prime_{\m\n\rho},h^\prime_{\m\ha}|
h_{\m\n},B_{\m\n\rho},h_{\m\ha}\ra=
h_{\m\n}^\prime h_{\m\n}+B^\prime_{\m\n\rho}B_{\m\n\rho}+
h_{\m\ha}^\prime h_{\m\ha}.\label{noname}
\ee
The $SO(9)$ and supersymmetric
invariance\footnote{Of course this is manifest when one inserts
$1=e^{-\L Q}e^{\L Q}$ into the left hand side of\rf{noname},
however this induces supersymmetry
transformations of the primed
polarisation tensors whose signs require some care.} 
of this result provides a simple check of our computations.

\sect{Construction of asymptotic states.}
\label{section3}

We now turn to the constrained $SU(2)$ sector of the theory describing
the relative motion of the two particles.
Similar to its $U(1)$ counterpart the state to be constructed
must be parametrised in terms of $SO(9)$ momentum and
polarisation tensors. Its construction, however, is complicated by
the constraint condition\rf{constraint}. Incidently, this condition even 
forbids
momentum eigenstates with nonzero eigenvalue. This follows from
the eigenstate equation $P^A_\m\, |p_\m\ra = p_\m \, |p_\m\ra$ 
(for some fixed $A$).
Acting twice with appropriate components of $\vec{L}$ 
and taking into account that
$\vec{L}\, |p_\m\ra=0$, along with the constraint algebra\rf{constraint},
implies that $p_\m$ vanishes. How can we then  hope
to find free particle states of definite momentum? The solution lies
in the fact that the states to be constructed are {\it not} exact momentum
eigenstates. Rather, they need only be eigenstates in the limit 
that the separation between the two particles
becomes large. Similarly they
will only transform correctly under the transverse Lorentz group $SO(9)$
in this asymptotic sense. These features fit in nicely with the advocated
principle of $M$(atrix) theory \cite{BFSS96} that the
concept of (commuting) spacetime emerges only in an asymptotic limit 
of the theory.

To find these asymptotic states let us suppose that the particles are
separated in (say) the ninth spatial direction. We have found it
useful to partially gauge fix the $SU(2)$ rotational invariance of 
the state by choosing a frame in which $\vec{X}_9$ lies
along the $z$-axis, i.e.\ $X^1_9=X^2_9=0$. 
We emphasize, however, that this is only a technical manouvre and all
our results may be re-expressed in terms of gauge invariant 
wavefunctions of the original variables.
We define
\bea
\vec{X}_9&=&(0,0,x) \nn \\
\vec{X}_a&=&(Y^1_a,Y^2_a,x_a)  \qquad \mbox{where }\  a=1,\ldots, 8.
\label{frame}
\eea
The Cartan subalgebra variables $x_\m=(x_a,x)$ will acquire the asymptotic
interpretation of the nine dimensional
spatial coordinates, whereas the $Y^I_a$ (where
$I=1,2$) describe the oscillatory ``off diagonal'' modes. Similarly
the fermions split up into $\theta^3$ taking care of the polarisation
structure of our state, whereas the $\theta^I$ serve
as the fermionic partners to the bosonic oscillators $Y^I_a$.
The following identification of supersymmetric harmonic oscillator
contributions to the $M$(atrix) theory Hamiltonian has already been employed
in the beautiful work of de Wit, L\"uscher and Nicolai \cite{dWLN89} to
prove the continuity of the supermembrane spectrum. 

To rewrite the Hamiltonian in the frame $Y^I\equiv X^I_9=0$,
derivatives with respect to $Y^I$ acting on
gauge invariant wavefunctions
must be reexpressed as
\bea
[\, \partial_{Y^I}\, ]_{Y^I=0}
&=& - \frac{i}{x}\, \epsilon^{IJ}\, \widehat{L}^J \nn \\
\left [ {\partial_x}^2+ {\partial_{Y^I}}^2 \right ]_{Y^I=0} &=&
\frac{1}{x}\, {\partial_x}^2\, x - \frac{1}{x^2}\, [ \, (\widehat{L}^1)^2+
(\widehat{L}^2)^2\, ]
\eea
where 
\be
\widehat{\hspace{.3mm} \vec{{L}}\hspace{.5mm} }= 
\vec{L} - \vec{X}_9\times \vec{P}_9
\ee
does not depend on $x$ or $Y^I$. Note however that\rf{frame} does
not completely fix the gauge, further $U(1)$ rotations about the
$z$-axis generated by the Cartan generator $L^3$ are still possible.

The Hamiltonian in this gauge then reads \cite{dWLN89}
\bea
H_{\rm V}&=& -\frac{1}{2x}\, (\dx)^2x -\frac{1}{2}\, (\dxa)^2 \label{H_V}\\
H_{\rm B}&=& -\frac{1}{2}\, (\frac{\partial}{\partial {Y_a^I}})^2
+ \frac{1}{2}\, r^2\, Y_a^I\, Y_a^I \label{H_B}\\
H_{\rm F}&=& r\, \tilde{\theta}^\dag\, \gam_9\, \tilde{\theta} \label{H_F}\\
H_4&=& \frac{1}{4}\, \epsilon^{IJ}Y^I_aY^J_b\epsilon^{KL}Y^K_aY^L_b 
- \frac{1}{2}\, x_a\, x_b\, Y_a^I
\, Y_b^I \nn \\ &&
+\frac{1}{x^2}\, (x_a\, \dxa +\frac{1}{2} \, Y_a^I\, 
\partial_{Y_a^I})
- \frac{1}{2\, x^2}\,x_a\, x_b \partial_{Y_a^I}\, \partial_{Y_b^I}\nn\\&&
- \frac{1}{2\, x^2}Y_a^I\, Y_b^I\, \dxa \partial_{x_b}
+\frac{1}{x^2}\, x_b\, Y_a^I\,\dxa \, \partial_{Y_b^I}
+ i\, \eps^{IJ}\, Y_a^I\, \theta^J\, \gam_a\, \theta^3\nn\\&& 
-\frac{1}{2\, x^2}\, (\theta^I\, \theta^3)\,
(\theta^I\, \theta^3) -\frac{1}{x^2} 
(\theta^I\, \theta^3)\, [ Y_a^I\,\dxa- x_a\,
\partial_{Y^I_a} ] \label{pretty}
\eea
where $r^2= x^2 +  x_a\, x_a$ and the spinors $\theta^I$ enter $H_{\rm F}$
through the complexified and spin($9$) rotated combination
\bea
\theta&=& \frac{1}{\sqrt{2}}(\theta^1+ i\, \theta^2) \label{comp}\\
\tilde{\theta}&=& \frac{r+\gam_9\gam_\mu x^\mu}{\sqrt{2r(r+x)}}\, \theta.
\label{rotissary}
\eea
These complex eight component 
spinors obey the canonical anti-commutation relations
\be
\{\, \theta_\a , \theta^\dagger_\b\, \} = 
\{\, \tilde{\theta}_\a , \tilde{\theta}^\dagger_\b\, \} = \delta_{\a\b},
\ee
although the $\tilde{\theta}_\a$ no longer commute with the bosonic
momentum operators.
Observe that the Hamiltonian~(\ref{H_V})-(\ref{pretty}) commutes with
the generator of residual $U(1)$ Cartan rotations
\be
L^3 = -i\, \epsilon^{IJ}\, Y^I_a\, \frac{\partial}{\partial\, Y^J_a}
- \frac{i}{2}\, \epsilon^{IJ}\, \theta^I_\ha\, \theta^J_\ha\, . \label{L_3}
\ee
Physical wavefunctions must, of course, be annihilated by $L^3$.


The sum of $H_{\rm B}$ and $H_{\rm F}$ in\rf{H_B} and\rf{H_F}, 
respectively, is the Hamiltonian of a
superharmonic oscillator in the transverse coordinates
$Y^I_a$ and $\theta^I_\ha$
with frequency $r$. The bosonic groundstate $\OB$ of\rf{H_B} is 
\bea
\OB&=& (r/\pi)^4\, e^{-\frac{r}{2}\, Y^I_a\, Y^I_a}\label{OB}\\
H_{\rm B}\, \OB &=& 8r\, \OB\ .\eea 
In the fermionic sector
we introduce the chiral spinors
\be
\gamma_9\, \tilde{\theta}_\pm = \pm \tilde{\theta}_\pm
\ee
to rewrite $H_{\rm F}$ as
\be
H_{\rm F}= r\, [\, ({\tilde{\theta}^\dagger}_+)_\a\, 
({\tilde{\theta}}_+)_\a + ({\tilde{\theta}}_{-})_\a\,
({\tilde{\theta}^\dagger}_-)_\a - 8\, ].
\ee
The fermionic groundstate is
\bea
\OF&=& \prod_{\a=1}^8\, (\ttheta^\dag_-)_\a\, |0_{\rm F(ock)}\rangle
\label{OF} \\
H_{\rm F}\, \OF&=& -8r \, \OF
\eea
where the canonical fermion vacuum $|0_{\rm F(ock)}\rangle$ is defined by
$\theta_\ha\,|0_{\rm F(ock)}\rangle=0$. 
The zero point energies of the combined system $H_{\rm B}+ H_{\rm F}$
cancel identically, as expected by supersymmetry. One may check~\cite{dWLN89} 
that the
groundstates\rf{OB} and\rf{OF} are annihilated by the residual 
constraint\rf{L_3}. We have normalised the states $\OB$ and $\OF$ to unity.

Let us now perform the split of the total Hamiltonian into a free
and interacting part. At first sight one might naively think that the
free Hamiltonian $H_0$ is given by $H_{\rm V}+H_{\rm B}+ H_{\rm F}$.
However, $H_{\rm V}$ is not diagonal in the superoscillator
space, as the frequency of the oscillator states depends 
on the Cartan variables $x_\m=(x_a,x)$.

We overcome this difficulty as follows. We would like to factor the Hilbert 
space into a product of transverse oscillatory degrees of freedom with
the space of particle-like excitations in the valley/flat directions. 
To this end one may envisage the $SU(2)$ Hilbert
space of our problem as a product (albeit non-direct) 
between the space  $\cal H_{\rm S}$ spanned
by all possible superoscillator states (denoted symbolically by $\{|n\ra\}$)
and the Cartan 
Hilbert space $\cal H_{\rm C}$
of wavefunctions depending on $x_\m=(x_a,x)$ and $\theta^3_\ha$. 
More concretely we write the total Hamiltonian as
\be
H=\sum_{n,m}\, |n\rangle\, \langle n|\, H \, |m\rangle\, \langle m|
\qquad \mbox{where}\quad\  \langle n|\, H \, |m\rangle \, : \, {\cal H_{\rm C}}
\rightarrow \, {\cal H_{\rm C} }\, ,
\ee
via the identity 
$\uni_{\cal H_{\rm SU(2)}}
=\sum_n|n\rangle\, \langle n|\times \uni_{\cal H_{\rm C}}$ with
$\uni_{\cal H_{\rm C}}=1$. Let us stress once more that the differential
operator $H_{nm}=\langle n|\, H \, |m\rangle$
does not commute with the states $|m\ra$.
It represents an ``effective'' Hamiltonian of the Cartan degrees of 
freedom pertaining to asymptotic spacetime. 

We may now identify  the free Hamiltonian
as the diagonal piece
\be
H_0=\sum_n \, |n\rangle\, \langle n|\left( H_{\rm V}+H_{\rm B}+H_{\rm F}
\, -\,\frac{c_n}{r^2}\right) |n\rangle\, 
\langle n|. \label{deserves}
\ee
The term $-c_n/r^2$ on the right hand side deserves some explanation.
Our scheme is that all terms in the Hamiltonian of order  $1/x$ should be 
relegated to the interaction part of the Hamiltonian. However, when one 
computes the expectations of the kinetic terms $H_{\rm V}$ between states 
$\langle n|$ and $|n\rangle$ one finds terms proportional to $1/r^2$ which we 
subtract off order by order as indicated.
Note that $c_0=9$. The ``effective'' Hamiltonian of $H_0$ 
acting in ${\cal H_{\rm C}}$
then takes the form
\be
(H_0)_{nn}= 
-\frac{1}{2x}\, (\dx)^2x -\frac{1}{2}\, (\dxa)^2 + d_n\, r\, .
\label{H_0_nn}
\ee
Crucially, $d_0=0$ as demonstrated above. Therefore we now have an obvious
candidate for the relative asymptotic supergraviton states
\be
|k_\m;h_{\m\n},B_{\m\n\rho},h_{\m\ha}\ra=\frac{1}{x}e^{ik_\m x_\m}
|h_{\m\n},B_{\m\n\rho},h_{\m\ha}\ra\otimes|0_B,0_F\ra\label{solutions2}.
\ee
Here, the supergraviton polarisation multiplet
$|h_{\m\n},B_{\m\n\rho},h_{\m\ha}\ra$ is built from the
$\underline{44}\oplus\underline{84}\oplus\underline{128}$  
representation of $\theta^3_\ha$ in complete analogy to the $U(1)$ sector
discussed in the previous section. Indeed, in accordance with the $M$(atrix)
theory picture in which the diagonal blocks describe individual
particle degrees
of freedom, one should perform this construction in terms of the variables
$\theta^0\pm\theta^3$ in order to obtain polarisations corresponding to
individual particle states.

The states\rf{solutions2} are true eigenstates of the free Hamiltonian
satisfying the correct free particle dispersion relation
\be
H_0\, |k_\m;h_{\m\n},B_{\m\n\rho},h_{\m\ha}\ra
=\frac{1}{2}k_\m k_\m\, |k_\m;h_{\m\n},B_{\m\n\rho},h_{\m\ha}\ra\ .
\ee
They are invariant under the residual gauge transformations\rf{L_3} 
as $L^3$ only acts on the $U(1)$ invariant
superharmonic oscillator vacuum $\OR$.
Furthermore they are plane wave normalisable\footnote{Note that the
measure for the $x$ coordinate is $4\pi\, x^2$ as is evident
from the gauge 
fixing procedure.}.

It follows from the outlined construction that the interaction
Hamiltonian $H_{\ss \INT}=H-H_0$ reads
\be
H_{\ss \INT}=
\sum_{n\neq m}|n\rangle\, \langle n|\, H_{\rm V}\, |m\rangle\,
\langle m|
+ \sum_{n,m}\, |n\rangle\, \langle n|\left(
\, H_4 + \frac{c_n}{r^2}\,\right) |m\rangle\, 
\langle m|.\label{quattro}
\ee
In the limit of large particle separation, $x\rightarrow\infty$, the
interaction Hamiltonian scales as
\be
\lim_{x\rightarrow \infty}H_{\ss \INT}= {\cal O}(1/\sqrt{x}).
\ee
Hence the interactions die off and the proposed supergraviton 
states\rf{solutions2} are seen to be free in this limit.
Let us explain this important observation in detail.
Expectation values of the operators $\theta^I$ between superoscillator
states $\la m|$ and $|n\ra$ do not scale with $x$.
However, it is well known that harmonic oscillator
expectations of $Y^I_a$ scale as the inverse square root of the
frequency. I.e., $\la m| Y^p |n\ra\sim r^{-p/2}$, which one can easily
see writing the $Y^I_a$ in terms of creation and annihilation operators
(see\rf{flippant}).

Similarly, our states are asymptotic momentum eigenstates, that
is, defining the spacetime momentum operator by \be{\cal P}_\m
\equiv -i
(\partial_x,\partial_{x_a})=
-i|\vec{X}_9|^{-1}\vec{X}_9\cdot\partial/\partial \vec{X}_\m\, ,\ee we have
\be
{\cal P}_\m\, |k_\m;h_{\m\n},B_{\m\n\rho},h_{\m\ha}\ra = k_\m\, 
|k_\m;h_{\m\n},B_{\m\n\rho},h_{\m\ha}\ra + {\cal O}(1/x)
\ee
as promised in the beginning of this section.

Let us remark at this point that the free Hamiltonian $H_0$ may also
possess normalisable eigenstates built from excited superoscillators of
the symbolic form $\psi(x_\m, \theta^3)\, |n\rangle$. However, their
energy will rise linearly with particle separation $x$ 
(due to the third term on the right hand side of\rf{H_0_nn}) 
and hence they do not qualify as scattering states.

What are the transformation properties of the states\rf{solutions2} 
under SO($9$)? The
SO($8$) subgroup is manifest in the index $a$, but the 
gauge fixing $Y^I=0$ of\rf{frame} breaks SO($9$). To gain some insight, 
let us study the state\rf{solutions2} in gauge unfixed
language. (This is a familiar problem, given some expression 
corresponding to some $SU(2)$ invariant quantity evaluated in a special
frame,
find the original manifestly invariant expression from which it came).
For example, the exponential term in\rf{solutions2} becomes
\be
(\vec{X}_9\cdot\vec{X}_9)^{-1/2}
\exp\, [i ( k\, \sqrt{\vec{X}_9\cdot\vec{X}_9}+ k_a\, 
\frac{\vec{X}_a\cdot\vec{X}_9}{\sqrt{\vec{X}_9\cdot\vec{X}_9}})]
\label{giexp}
\ee
which is manifestly gauge invariant 
as it is built only from dot products of $SU(2)$ 3-vectors.
The above expression reverts to  $x^{-1}\exp{i(k\,x\ + k_a\, x_a)}$
in the $Y_I=0$ gauge. Similarly the term $r\, Y^I_a\, Y^I_a$ 
appearing in the exponential of $\OB$ of\rf{OB} may be 
written gauge invariantly as 
\be
\left(\vec{X}_9\cdot\vec{X}_9 + \frac{(\vec{X}_a\cdot\vec{X}_9)^2}
{\vec{X}_9\cdot\vec{X}_9}\right)^{1/2}
\, \frac{(\vec{X}_9 \times \vec{X}_a)^2}
{\vec{X}_9\cdot\vec{X}_9}.\label{gi2}
\ee
One may make analogous calculations for the fermions.
In this gauge unfixed language the SO($9$) Lorentz transformations are
obvious, e.g.\ an infinitesimal rotation $\epsilon$ in the 8-9 plane reads
\bea
\vec{X}_9 &\rightarrow& \vec{X}_9 + \epsilon
\, \vec{X}_8 \\
\vec{X}_8 &\rightarrow& \vec{X}_8 - \epsilon\,  
\vec{X}_9\, ,
\eea
under which\rf{giexp} and\rf{gi2} transform non-covariantly.
However, in the limit $|\vec{X}_9|=\sqrt{\vec{X}_9\cdot\vec{X}_9}
\rightarrow \infty$ all contributions violating $SO(9)$ 
covariance are suppressed. A similar
observation holds for supersymmetry. This blends nicely with the advocated
principle of $M$(atrix) theory in which the spacetime interpretation arises
only in the asymptotic regime of widely separated particles. 
Hence we see that our states\rf{solutions2} possess all the
requirements for free particle states in the 
limit of large separation, i.e.\ $SO(9)$
covariance, the correct dispersion relations, 
definite momenta and polarisations.

\sect{The Born approximation (and beyond).}
\label{section4}
The leading term in conventional quantum mechanical scattering theory is 
the Born amplitude and its computation is the subject of this section. 
However, we shall find a rather different type of expansion
to that encountered in a loopwise diagrammatical expansion of the
supersymmetric Yang-Mills theory~\cite{BB97,BBPT97,BM97}. To see this let
us reinstate the factors $M$ and $R$ as in\rf{ham} and
make the following rescalings 
$\vec{X}_\m\rar M^{-3}\vec{X}_\m$, $\vec{P}_\m\rar M^{-3}\vec{P}_\m$
and $\vec{\theta}\rar M^{-3}\vec{\theta}$. One then finds
\be
H=\frac{R}{M^6}\H
\ee
where $$\H=\frac{1}{2}\vec{P}_\m\cdot\vec{P}_\m
+\frac{1}{4}(\vec{X}_\m\times\vec{X}_\m)^2+
\frac{i}{2}\vec{X}_\m\cdot
\vec{\theta}\g_\m\times\vec{\theta}$$ and the canonical commutation
relations now read
\bea
[X^A_\m,P^B_\n]&=&iM^6\delta_{\m\n}\delta^{AB}\\
\{\theta^A_\ha,\theta^B_\hb\}&=&M^6\delta_{\ha\hb}\delta^{AB}\, .
\eea
We see that $M^6\equiv\hbar$ plays the r\^ole of $\hbar$ and henceforth
will be denoted as such. A loopwise expansion leads, of course, to an
expansion in $\hbar$, but this will not be the case for our quantum 
mechanical expansion as we shall see from the following developments.

Employing the results of the previous sections the computation of a
$2\rightarrow 2$ supergraviton amplitude in $M$(atrix) theory
may now be performed via elementary quantum mechanical scattering
theory. The centre of mass part is trivial and the relative
piece of the Born amplitude, using\rf{quattro}, reads
\bea
\la  
k^\pr_\m;h^\pr_{\m\n},B^\pr_{\m\n\rho},h^\pr_{\m\ha}|\, H_{\rm Int}\, |
k_\m;h_{\m\n},B_{\m\n\rho},h_{\m\ha}\ra&&\nn\\
&\hspace{-6.4cm}=\hspace{.2cm}\int{4\pi x^2 d^9x}\, 
\frac{\textstyle e^{ik_\m^\prime x_\m}}{\textstyle x}\,
\la h^\pr_{\m\n},B^\pr_{\m\n\rho},h^\pr_{\m\ha}|\, H_{\rm  
Eff}(x_\m,\partial_\m,
\theta^3_\ha)\,
|h_{\m\n},B_{\m\n\rho},h_{\m\ha}\ra 
\, \frac{\textstyle e^{ik_\m x_\m}}{\textstyle x}
&\nn
\eea\be\ee
where
\be
H_{\rm Eff}(x_\m,\partial_\m,\theta^3_\ha)\equiv
\la0_B,0_F|H_{\rm Int}|0_B,0_F\ra.
\ee
Therefore the $M$(atrix) theory Born amplitude takes the form of a
conventional nine-dimensional Born amplitude, namely the nine dimensional
Fourier transform of some interaction Hamiltonian $H_{\rm Eff}$. The new
ingredient is that the ``effective'' interaction Hamiltonian  
$H_{\rm Eff}$ is obtained as the superoscillator vacuum
expectation value of the full interaction Hamiltonian $H_{\rm Int}$.
Let us now turn our attention to the calculation of this vacuum  
expectation.

In rescaled variables, the object we wish to calculate is
\be
\lbf (\H_4+\frac{c_0}{r^2} )\rbf\ ,\label{object}
\ee
where
\bea
\H_4
&=& \frac{1}{4}\, \epsilon^{IJ}Y^I_aY^J_b\epsilon^{KL}Y^K_aY^L_b
- \frac{1}{2}\, x_a\, x_b\, Y_a^I
\, Y_b^I \nn
\\ &&
+\frac{\hbar^2}{x^2}\, (x_a\, \dxa +\frac{1}{2} \, Y_a^I\,
\partial_{Y_a^I})
- \frac{\hbar^2}{2\, x^2}\,x_a\, x_b \partial_{Y_a^I}\,  
\partial_{Y_b^I}\nn\\&&
- \frac{\hbar^2}{2\, x^2}Y_a^I\, Y_b^I\, \dxa \partial_{x_b}
+\frac{\hbar^2}{x^2}\, x_b\, Y_a^I\,\dxa \, \partial_{Y_b^I}
+ i\, \eps^{IJ}\, Y_a^I\, \theta^J\, \gam_a\, \theta^3\nn\\&&
-\frac{1}{2\, x^2}\, (\theta^I\, \theta^3)\,
(\theta^I\, \theta^3) -\frac{\hbar}{x^2}
(\theta^I\, \theta^3)\, [ Y_a^I\,\dxa- x_a\,
\partial_{Y^I_a} ] \label{petty}
\eea
and, as explained in section~\ref{section3}, the constant $c_0$ is  
fixed such that
\be
\lbf\Bigl(-\frac{\hbar^2}{2x}\partial_{x_\m}^2x-\frac{c_0}{r^2}\Bigr)\rbf=
-\frac{\hbar^2}{2 x}\partial_{x_\m}^2x\ .
\ee
To handle the non-commutativity of derivatives
$\partial_{x_\m}$ with superoscillator states we define
mode operators
\bea
Y_a^I= \frac{ a_a^I+{a_a^I}^\dag}{\sqrt{2r/\hbar}}&,&
\partial_{Y_a^I}= \frac{a_a^I-{a_a^I}^\dag}{\sqrt{2\hbar/r}}\,  ,
\label{mode}\label{flippant}\\
{}[a^I_a,{a^J_b}^\dagger]&=&\delta_{ab}\, \delta^{IJ}\, ,
\eea
whose derivatives read
\bea
\partial_{x^\mu}\, a^I_a= \frac{1}{2}\, \frac{x^\mu}{r^2}\,
(a^I_a)^\dag &,&
\partial_{x^\mu}\, (a^I_a)^\dag= \frac{1}{2}\, \frac{x^\mu}{r^2}\,
a^I_a\, .
\eea
Derivatives with respect to $x_\m$ on bosonic oscillator  
wavefunctions can be equivalently
expressed in terms of combinations of these mode operators acting on states.
The $x_\m$ dependence of the
spin(9) rotated spinor variables $\tilde{\theta}$ is given
by\rf{rotissary} so that the fermion oscillator
vacuum state is also $x_\m$ dependent. Using these results
one finds the following for derivatives on the superoscillator
ground state
\bea
\left[\partial_{x_\m},\rbf\right]
&=&\left(
- \frac{x^\mu}{4\, r^2}\,  {a_a^I}^\dag\,  {a_a^I}^\dag\right. \nn\\&&
\!\!\!\!\!\! +\frac{1}{2}\left. \left[
(\delta_{\m9}+\frac{(1-\delta_{\m9})x_\m}{r+x})
\frac{x_a}{r^2}(\tilde{\theta}^\dagger\g_a\tilde{\theta})
-\frac{1-\delta_{\m9}}{r}(\tilde{\theta}^\dagger\g_\m\tilde{\theta})
\right]
\right)\rbf\nn\\&&
\eea
where no summation is implied over $\m$.
Higher derivatives can be computed in a similar fashion, in particular,
the vacuum expectation value of the Cartan kinetic operator is
\be
\lbf\, \frac{-\hbar^2}{2x}\partial_{x_\m}^2x\, \rbf=
-\frac{\hbar^2}{2x}\, \partial_{x_\m}^2\, x+\frac{9\hbar^2}{r^2} .
\ee
yielding, as promised, $c_0=9\hbar^2$.

Using the definitions of the mode operators\rf{mode} and the above  
formulae
for derivatives on the superoscillator groundstate  plus 
analogous formulas for the 
fermions, the computation  
of\rf{object} is reduced to some simple mode operator algebra and  
we find
\bea
\!\!\!\!\lbf \, (\H_4+\frac{c_0}{r^2} )\, \rbf&=& \frac{16\hbar^2}{r^2}  
+\frac{\hbar(r^2-x^2)^2}{2\, x^2\, r}
-\frac{\hbar^2(r^2-x^2)}{2\, x^2\, r^2} \nn\\&&
 +\frac{\hbar^3(19r^2+x^2)}{2x^2\, r^5}
+\frac{\hbar^3 x_a}{2\, x^2\, r^3}\, \dxa-\frac{\hbar^3}{2\, x^2\,  
r}\, (\dxa)^2.\nn\\&&\label{Born}
\eea
In terms of an expansion in large $x$, the leading term, $16/r^2$,
is $SO(9)$ invariant in accordance with our previous arguments that
$SO(9)$ symmetry should hold to leading order in a large $x$ expansion.
Furthermore there is no explicit $\theta^3$ dependence which matches 
the suspicion that spin effects should be found only at higher orders.
However, the quantum mechanical perturbation theory in which transverse
oscillator modes are effectively ``integrated out'' is not organised
directly as an expansion in $\hbar$. 

The perturbative expansion we are performing is an expansion in
$1/x$ of the effective potential. However, as the Born 
approximation\rf{Born} shows, this 
expansion does not scale uniformly in $1/x$ for every given
order, rather there will be an overlap of contributions to
a specific $1/x^n$ term from a finite number of orders in
quantum mechanical perturbation theory. In particular the
leading $16\hbar/r^2$ term of\rf{Born} also receives contributions in
second order perturbation theory, but there will be no further
contributions at higher orders as a simple scaling analysis of
the interaction terms in\rf{petty} shows. We hence proceed to evaluate
this
contribution. 

Here fortune smiles upon us since the sole contribution to order $r^{-2}$
in the effective potential at second order perturbation theory stems from 
\be
V_2=\lbf (i\epsilon^{KL}\, Y^K_b\, \theta^L\, \gamma_b\, \theta^3)\,
\frac{1}{E-H_0+i\epsilon}\, (i\epsilon^{IJ}\, Y^I_a\, \, \theta^J
\gamma_a\, \theta^3)\rbf\, .
\label{2ndorder}
\ee
To see this observe that $Y^I_a$ scales
as $r^{-1/2}$ from\rf{mode} and the leading 
contribution to $H_0$ is linear in $r$ as\rf{H_0_nn} shows. 
As a matter of fact 
this linear contribution to $H_0$ is the only one entering the $r^{-2}$
part of $V_2$ and hence
\bea
V_2&=&\sum_n \lbf(\epsilon^{KL}\, Y^K_b\, \theta^3\, \gamma_b\, \theta^L)
|n\rangle\, \langle n| \frac{-1}{d_n \, \hbar\, r}|n\rangle\, \langle n|
\, (\epsilon^{IJ}\, Y^I_a\, \theta^3\, \gamma_a\, \theta^J)\rbf
\nn\\&& \hspace{9.5cm}+ \, {\cal O}(1/x^3)
\eea
Now reading this amplitude
from the right hand side, we see that the interaction term 
homogeneously excites one bosonic and one fermionic oscillator mode.
Thus the sum over 
oscillator projectors collapses to the $n=2$ sector with $d_2=2$
and we are left with the computation of a norm.
Therefore, dropping the ${\cal O}(1/x^3)$ terms, we have
\bea
V_2= -\frac{1}{2\, \hbar\, r}\, \left|\left|\epsilon^{IJ}\, Y^I_a\, 
\theta^J\, \gamma_a\, \theta^3\rbf\right|\right|^2 
= -\frac{16\, \hbar^2}{r^2}
\eea
which one shows using\rf{flippant},  
$\lbf \theta^I_\ha\, \theta^I_\hb\rbf=\delta_{\ha\hb}$
and $(\theta^3_\ha\, \theta^3_\ha)=8$ due to the canonical anti-commutation
relations.

But this contribution cancels exactly with the one obtained in 
the Born amplitude\rf{Born}. Therefore the overall $1/r^2$ dependence
of the effective potential for two widely separated supergravitons
vanishes! 
This result agrees with the effective action found in a path integral 
background field calculation~\cite{BB97} where a possible 
$\hbar/r^2$ contribution arises at two loops but also vanishes.
A priori the connection
between this field theoretic effective action and our canonical effective
Hamiltonian for the Cartan degrees of freedom is not at all obvious.
Nonetheless, this supersymmetric cancellation is a strong confirmation of
our proposal.
Furthermore, the understanding of
the precise connection between the two approaches that we provide, 
yields an effective and transparent means of computing
the $M$(atrix) theory $S$ matrix for eleven dimensional supergravitons.

As demonstrated in \cite{BBPT97} the loopwise expansion of the
path integral background field effective action is an expansion in
$\hbar\equiv M^6$. However our quantum mechanical perturbation theory
is manifestly not organised as an expansion in $\hbar$, e.g.,\ the Born
amplitude\rf{Born} is comprised of terms arising from one, two and
three loop level from the point of view of \cite{BBPT97}. In fact every
order in quantum mechanical perturbation theory will give contributions
to a given order of $\hbar$. One may see this by studying the $\hbar$
dependence
of the $m$-th order perturbation theory amplitude
\be
\lbf H_{\rm Int}\, \frac{1}{E-H_0+i\epsilon}\, H_{\rm Int}\,
\frac{1}{E-H_0+i\epsilon}\, H_{\rm Int}\, \ldots 
\frac{1}{E-H_0+i\epsilon}\, H_{\rm Int}\rbf
\ee
Now as $H_{\rm Int}\approx(1+\hbar+\hbar^2)$ and
\mbox{$E-H_0\approx (1+\hbar)$} we see that at any order of $\hbar$
all orders of quantum mechanical perturbation theory contribute. 
Let us stress again, that this conclusion applies 
only to the $\hbar$ dependence of the effective Hamiltonian. The
expansion in $1/x$ is well under control in the sense that for
a given order in $1/x$ we will only have to compute a finite number
of orders in perturbation theory, just as was the case for 
the $1/r^2$ term in the above.


\sect{Conclusion.}
\label{section6}

Although we have not yet carried out the goal of computing the 
$M$(atrix) theory supergraviton $S$ matrix it is clear that our
proposal provides a transparent framework for such a
calculation. We have also reached the point at which one can
pause and consider how this calculation can be most efficiently undertaken.
Looking back however, it is pleasing to see that the $M$(atrix) theory
really admits 
asymptotic
quantum mechanical wavefunctions with the correct interpretation
as supergraviton particle states possessing the appropriate polarisations,
momenta and dispersion relations.

Certainly it is possible to carry out second and higher order 
quantum mechanical perturbation theory in the manner indicated in 
section~\ref{section4}. Such a calculation, despite its somewhat
untidy appearance, is not inordinately difficult and yields
{\it directly} $M$(atrix) theory $S$ matrix elements. This is to be contrasted
with the existing background path integral approaches relevant to 
eikonal scattering from which one extracts phase shifts and in turn
the effective potential for a $D0$ brane probe moving under the
influence of a $D0$ brane source. Ultimately, our results should
be directly comparable with $S$ matrix amplitudes of eleven dimensional 
supergravity, rather than simply the effective potential for a probe
moving in a classical eleven dimensional gravitational background.
Furthermore, the manifest inclusion of spin degrees of freedom
through the physical polarisation tensors of the eleven dimensional
supergraviton multiplet allows spin effects to be directly 
calculated\footnote{The first $M$(atrix) theory
calculations of the spin-orbit interactions
of a spinning $D0$ brane probe have recently been 
carried out in~\cite{K97}.}.
 
Note that our calculation of the $M$(atrix) theory
Born amplitude considered the case in which the large ingoing and 
outgoing separations of the supergraviton particle pairs were both in the
nine direction which is the kinematical regime described by the 
eikonal approximation. There is however, in principle, no impediment to 
considering more general kinematical situations (see the appendix
for further discussion of this point). 

Nonetheless, efficient perturbative calculations should also be possible
by a loopwise computation of the generating functional of one particle
irreducible Greens functions (the effective action), 
which is readily obtainable 
from existing background field computations evaluated at arbitrary
values of the background fields. To this end one needs 
reduction formulae connecting these Greens functions with the
scattering amplitudes of our asymptotic states. However, for a first 
quantized quantum mechanical model such as the $M$(atrix) theory
the solution to this problem is well known (for completeness,
a brief derivation of the following formulae is 
presented in the appendix). 
The result for the
relevant $S$ matrix elements is given by
\be
S_{fi}=\int d^9\vec{X}^\prime\, d^{16}\vec{\theta}^\prime\,
              d^9\vec{X}\, d^{16}\vec{\theta}\,\,
\Phi^*_f(\vec{X}_\m^\prime,\vec{\theta}_\ha^\prime)\,
\la \vec{X}_\m^\prime,\vec{\theta}_\ha^\prime|\, U(\infty,-\infty)\,
|\vec{X}_\m,\vec{\theta}_\ha\ra\,\,
\Phi_i(\vec{X}_\m,\vec{\theta}_\ha)
\ee
where we have rewritten our ingoing and outgoing
asymptotic supergraviton
states as wavefunctions $\Phi_i(\vec{X}_\m,\vec{\theta}_\ha)$
and $\Phi^*_f(\vec{X}_\m^\prime,\vec{\theta}_\ha^\prime)$ respectively.
The quantum mechanical transition element 
$\la \vec{X}_\m^\prime,\vec{\theta}_\ha^\prime|\, U(\infty,-\infty)\,
|\vec{X}_\m,\vec{\theta}_\ha\ra$
may be expressed as a gauge fixed BRST symmetric path integral in the standard
way
\bea
\la \vec{X}_\m^\prime,\vec{\theta}_\ha^\prime|\, U(\infty,-\infty)\,
|\vec{X}_\m,\vec{\theta}_\ha\ra&=&
\lim_{T\rightarrow\infty}
\int^{{}^{{}^{\vec{X}_\m(T/2)=\vec{X}_\m^\prime}_{
\vec{\theta}_\ha(T/2)=\vec{\theta}^\prime_\ha}}}_{{}_{
{}^{
\vec{X}_\m(-T/2)=\vec{X}_\m}_{
\vec{\theta}_\ha(-T/2)=\vec{\theta}_\ha}}}
D^9\vec{X}\,D^{16}\vec{\theta}\,D\vec{A}\,D\vec{b}\,D\vec{c}\,\,
e^{iS_{\rm BRST}}\, .\nn\\&&
\eea
The boundary conditions of this path integral
can be rendered standard by the introduction of 
appropriate background fields. The BRST action is given (in covariant gauge) 
by
\be
S_{\rm BRST}=S_{\rm SYM}+
\int_{-T/2}^{T/2} dt\left(\frac{1}{2}\dot{\vec{A}}^2+\dot{\vec{b}}
\cdot \frac{D(\vec{A})c}{Dt}\right),
\ee
where $S_{\rm SYM}$ is the usual one dimensional
super Yang-Mills action of the $M$(atrix) theory.
One is now in the realm of the existing path integral calculations
that one finds in the literature~\cite{DKPS97,Ael97,BB97,BBPT97} 
and the connection between these calculations
and supergraviton $S$ matrix elements is now clear.

Penultimately we mention that 
we have only considered asymptotic states of the $U(2)$
$M$(atrix) theory describing a pair of widely separated particles.
Three body graviton scattering, for example, would require knowledge
of the $SU(2)$ zero energy ground state wavefunction, since a single outgoing 
supergraviton state is described by the product of this ground state with
the centre of mass $U(1)$ supergraviton wave function outlined above.
Similar ground state considerations are necessary in the study of pairs
of particles in the $U(N)$ $M$(atrix) theory for $N>2$
where one could even describe an exchange of momentum in the $x_-$
direction. Existence properties of such ground states are discussed in
\cite{SS,PR97} and progress towards their
construction has recently been made in~\cite{H97}. 

Our final remark concerns the construction of $n$-body asymptotic states
for $n>2$. It is obvious how our scheme can be generalised to the limited
kinematical regime in which the particles are collinear. However,
more general kinematical situations require further input. Nonetheless,
such a generalisation should certainly draw its main ingredients from our
asymptotic state construction.

\sect{Acknowledgements.}

We would like to thank Ruben Minasian and Stefan Theisen for their
collaboration in the early stages of this work.
We also acknowledge useful discussions with Bernard de Wit, Robbert Dijkgraaf,
Marco Serone, Erik Verlinde and Herman Verlinde. One of us (A.W.)
would like to
thank Koenraad Schalm for lengthy discussions on quantum mechanical
gauge theories.
Finally, we thank our colleagues at NIKHEF, Peter Jarvis,
Kasper Peeters, Bert Schellekens and Jan Willem van Holten
for listening to
our questions from the left field. 

\pagebreak
\section*{Appendix}
\renewcommand{\theequation}{A.\arabic{equation}}
\setcounter{equation}{0}
\subsection*{A. Reduction formulae for $M$(atrix theory).}
\label{appendix}

In this appendix we outline the $M$(atrix)
theory analogues of the LSZ reduction formulae relating
Greens functions and $S$ matrix elements. 
Detailed discussions of the below considerations in a more general
context may be found in the review and books listed in~\cite{Abers}.

We wish to compute $S$ matrix elements of the form
\be
S_{fi}=
\la  
k^\pr_\m;h^\pr_{\m\n},B^\pr_{\m\n\rho},h^\pr_{\m\ha};\vec{X}_{9^\prime}^\prime
|\, U(\infty,-\infty)\, |
k_\m;h_{\m\n},B_{\m\n\rho},h_{\m\ha};\vec{X}_9\ra
\label{S}
\ee
with $U(\infty,-\infty)=\lim_{T\rightarrow\infty}\exp(-iHT)$. The in and out 
states in\rf{S} are our asymptotic supergraviton states written in a general, 
ungauge-fixed frame in variables $(\vec{X}_\m,\vec{\theta}_\ha)$.
The additional labels $\vec{X}_9$ and $\vec{X}_{9^\prime}^\prime$,
respectively, indicate the direction
in which the asymptotic ingoing and outgoing particles
are widely separated. Our earlier calculation of the
Born amplitude considered the case in which the large ingoing and 
outgoing separations of the supergraviton particle pairs were {\it both} in the
nine direction which is the kinematical regime described by the 
eikonal approximation. However, here we take the more
general kinematical situation in which 
$\vec{X}_9\neq\vec{X}_{9^\prime}^\prime$.

Introducing twice the resolution of unity in the form\footnote{Here 
we are somewhat schematic. For the sixteen dimensional real spinors 
$\vec{\theta}_\ha$ one should really complexify and build coherent states.
Technical problems of this type have been considered
in the context of quantum mechanical path 
integration in~\cite{Peeters}.}
\be\uni=\int d^9\vec{X}\, d^{16}\vec{\theta}\, |\vec{X}_\m,\vec{\theta}
_\ha\ra \, \la \vec{X}_\m,\vec{\theta}_\ha|\ee
and denoting our asymptotic in and out states in wavefunction form as
$\Phi_i(\vec{X}_\m,\vec{\theta}_\ha)$ and 
$\Phi_f(\vec{X}_\m,\vec{\theta}_\ha)$, respectively, we have 
\be
S_{fi}=\int d^9\vec{X}^\prime\, d^{16}\vec{\theta}^\prime\,
              d^9\vec{X}\, d^{16}\vec{\theta}\,\,
\Phi^*_f(\vec{X}_\m^\prime,\vec{\theta}_\ha^\prime)\,
\la \vec{X}_\m^\prime,\vec{\theta}_\ha^\prime|\, U(\infty,-\infty)\,
|\vec{X}_\m,\vec{\theta}_\ha\ra\,\,
\Phi_i(\vec{X}_\m,\vec{\theta}_\ha)
\ee
The quantum mechanical
transition element
$\la \vec{X}_\m^\prime,\vec{\theta}_\ha^\prime|\, U(\infty,-\infty)\,
|\vec{X}_\m,\vec{\theta}_\ha\ra$
may be expressed as a path integral in the usual way. If one is interested in
(say) covariant gauges it is necessary to introduce the Hermitean and
nilpotent BRST charge
\be
Q=\vec{c}\cdot\vec{L}+\vec{p}_b\cdot\vec{p}_A+\frac{1}{2}\,
(\vec{c}\times\vec{c})
\cdot\vec{p}_c
\ee
along with the ghosts $\vec{c}$ and antighosts $\vec{b}$ and their respective
canonical momenta $\vec{p}_c$ and $\vec{p}_b$. Furthermore, we 
reinstate the Lagrange multiplier $\vec{A}$ of the original super Yang-Mills
theory and its canonical momentum $\vec{p}_A$.
Our gauge invariant asymptotic states are clearly physical
and therefore annihilated by $Q$ without being exact
and so, according to the Batalin Vilkovisky theorem, 
we are free\footnote{In principal one must 
 be somewhat careful here, but
for technical details we refer the reader to the texts~\cite{Abers}.} 
to add a piece to the Hamiltonian
$\{Q,\Psi\}$ for some gauge fixing fermion $\Psi$.
In particular Feynman gauge is reached via the choice
\be
\Psi=i\vec{p}_c\cdot\vec{A}+\frac{i}{2}\vec{b}\cdot\vec{p}_A.
\ee
After integrating out appropriate canonical momenta via their algebraic 
field equations the gauge fixed path integral result for the 
transition element is
\bea
\la \vec{X}_\m^\prime,\vec{\theta}_\ha^\prime|\, U(\infty,-\infty)\,
|\vec{X}_\m,\vec{\theta}_\ha\ra&=&
\lim_{T\rightarrow\infty}
\int^{{}^{{}^{\vec{X}_\m(T/2)=\vec{X}_\m^\prime}_{
\vec{\theta}_\ha(T/2)=\vec{\theta}^\prime_\ha}}}_{{}_{
{}^{
\vec{X}_\m(-T/2)=\vec{X}_\m}_{
\vec{\theta}_\ha(-T/2)=\vec{\theta}_\ha}}}
D^9\vec{X}\,D^{16}\vec{\theta}\,D\vec{A}\,D\vec{b}\,D\vec{c}\,\,
e^{iS_{\rm BRST}}\nn\\&&\label{vorletzte}\\
&\equiv&e^{i\Gamma(\vec{X}^\prime_\m,\vec{\theta}_\ha^\prime;
\vec{X}_\m,\vec{\theta}_\ha)}
\label{letzte}
\eea
where the BRST action is the sum of the dimensionally reduced
one dimensional super Yang-Mills action and
gauge fixing terms $S_{\rm BRST}=S_{\rm SYM}+S_{\rm Fix}$
with
\be
S_{\rm Fix}=\int_{-T/2}^{T/2} dt\left(\frac{1}{2}\dot{\vec{A}}^2+\dot{\vec{b}}
\cdot \frac{D(\vec{A})c}{Dt}\right).
\ee
The perturbative
computation of $\Gamma(\vec{X}^\prime_\m,\vec{\theta}_\ha^\prime;
\vec{X}_\m,\vec{\theta}_\ha)$ may be carried out 
by a straightforward 
loopwise expansion of the path integral\rf{vorletzte}.



\newpage


\begin{thebibliography}{99}
\bibitem{BFSS96} T.\ Banks, W.\ Fischler, S.H.\ Shenker and L.\ Susskind, 
\PRD{55}{1997}{5112}, { hep-th/9610043}.
\bibitem{CH} M.\ Claudson and M.B.\ Halpern, Nucl.\ Phys.\ B250 (1985) 
689;\\
R.\ Flume, Ann.\ Phys.\ {164} (1985) 189; \\
M.\ Baake, P.\ Reinicke, and V.\ Rittenberg, J.\ Math.\
Phys.\ {26} (1985) 1070. 
\bibitem{Witten95} E.\ Witten, \NPB{443}{1995}{85}.
\bibitem{Susskind97} L.\ Susskind, ``Another Conjecture about
$M$(atrix) Theory'', { hep-th/9704080}.
\bibitem{DKPS97} M.D.\ Douglas, D.\ Kabat, P.\ Pouliot and S.\ Shenker,
\NPB{485}{1997}{85}, hep-th/9608024.
\bibitem{Ael97}
O.\ Aharony and M.\ Berkooz, \NPB{491}{1997}{184}, hep-th/9611215;\\
G.\ Lifschytz and S.D.\ Mathur, ``Supersymmetry and Membrane Interactions
in M(atrix) Theory'', hep-th/9612087;\\
G.\ Lifschytz, ``Four Brane and Six Brane Interactions in M(atrix) Theory'',
hep-th/9612223;\\
D.\ Berenstein and R.\ Corrado, \PLB{406}{1997}{37}, hep-th/9702108;\\
V.\ Balasubramanian and F.\ Larsen, ``Relativistic Brane Scattering'', 
hep-th/9703039.
\bibitem{BB97} K.\ Becker and  M.\ Becker, ``A Two-Loop Test of
$M$(atrix) Theory'', { hep-th/9705091}.
\bibitem{BBPT97} K.\ Becker, M.\ Becker, J.\ Polchinski and A.\
Tseytlin, \PRD{56}{1997}{3174}, { hep-th/9706072}.
\bibitem{FP67} L.D. Faddeev and V.N. Popov, Phys. Lett. B25 (1967) 29.
\bibitem{Background} B.S.\ DeWitt, Phys.\ Rev.\ 162 (1967) 1195;\\
 G.\ 't Hooft, Acta Universitatis Wratislavensis No.\ 38 XII, Winter School
of Theoretical Physics in Karpacz; ``Functional and Probabilistic Methods
in Quantum Field Theory'', Vol.\ 1 (1975);\\
L.F.\ Abbott, Acta Phys.\ Pol.\ B13 (1992) 33. 
\bibitem{LSZ59}H. Lehmann, K. Syzmanzik and W. Zimmermann,
Nuov. Cim. 1 (1955) 205.
\bibitem{dWHN88} B.\ de Wit, J.\ Hoppe and H.\ Nicolai, \NPB{305}{1988}{545}.
\bibitem{Polchinski95} J.\ Polchinski, Phys.\ Rev.\ Lett.\ 75 (1995) 4724,
hep-th/9510017.
\bibitem{Witten96} E.\ Witten, \NPB{460}{1996}{335}, hep-th/9510135.
\bibitem{D0-branes}U.\ Danielson, G.\ Ferretti and B.\ Sundborg, 
Int.\ J.\ Mod.\ Phys.\ A11 
(1996) 5463, { hep-th/9603081};\\
D.\ Kabat and P.\ Pouliot, 
Phys.\ Rev.\ Lett.\ 77 (1996) 1004, { hep-th/9603127}.
\bibitem{SS} S. Sethi and M. Stern, ``D-Brane Bound States Redux'',
{ hep-th/9705046}.
\bibitem{dWLN89} B.\ de Wit, M.\ L\"uscher and H.\ Nicolai, 
\NPB{320}{1989}{135}.
\bibitem{BM97}P. Berglund and D. Minic, ``A Note on Effective  
Lagrangians in Matrix Theory'', { hep-th/9708063}.
\bibitem{K97}P. Kraus, ``Spin-Orbit Interaction from Matrix Theory'',
{ hep-th/9709199}.
\bibitem{Abers}
E.S. Abers and B.W. Lee, Phys. Rep. 9 (1973) 1;\\
J. Govaerts, ``Hamiltonian Quantizations and Constrained
Dynamics'', Leuven notes in mathematical and theoretical physics, 
Leuven 1991;\\
M. Henneaux and C. Teitelboim, ``Quantization of Gauge Systems'',
Princeton University Press, Princeton 1992. 
\bibitem{PR97} M.\ Porrati and A.\ Rozenberg, ``Bound States at Threshold
in Supersymmetric Quantum Mechanics'', hep-th/9708119.
\bibitem{H97}J. Hoppe, 
``On the Construction of Zero Energy States in Supersymmetric Matrix Models'',
{ hep-th/9709132};
``On the Construction of Zero Energy States 
in Supersymmetric Matrix Models II'', { hep-th/9709217}.
\bibitem{Peeters} J. de Boer, B. Peeters, K. Skenderis and P. 
van Nieuwenhuizen, \NPB{459}{1996}{631}.
\end{thebibliography}
\end{document}